\documentstyle[emulateapj,psfig,pstricks,apjfonts]{article}
\newcommand{\Hii}{\ion{H}{2}}

\newcommand{\snr}{{G266.2$-$1.2}}
\newcommand{\rxj}{{RXJ~0852.0$-$0462}}

\begin{document}

\title{\rxj: Another Nonthermal Shell-Type SNR (\snr)}

\author{Patrick Slane\altaffilmark{1},
John P. Hughes\altaffilmark{2}, 
Richard J. Edgar\altaffilmark{1},
Paul P. Plucinsky\altaffilmark{1},\\ 
Emi Miyata\altaffilmark{3,4},
Hiroshi Tsunemi\altaffilmark{3,4},
and Bernd Aschenbach\altaffilmark{5}}

\altaffiltext{1}{Harvard-Smithsonian Center for Astrophysics, 60 Garden Street,
Cambridge, MA 02138}
\altaffiltext{2}{Department of Physics and Astronomy, Rutgers, The State 
University of New Jersey, 136 Frelinghuysen Road, Piscataway, NJ 08854-8019}
\altaffiltext{3}{Department of Earth and Space Science, Osaka University,
1-1 Machikaneyama, Toyonaka, Osaka 560-0043 JAPAN}
\altaffiltext{4}{CREST, Japan Science and Technology Corporation, 2-1-6 
Sengen, Tsukuba, Ibaraki 305-0047 JAPAN}
\altaffiltext{5}{Max-Planck-Institut f\"{u}r extraterrestrische Physik, 
D-85740, Garching, Germany}

\slugcomment{Accepted for publication in The Astrophysical Journal}

\begin{abstract}
\noindent
The newly discovered supernova remnant \snr\ (RX~J0852.0--0462), along the 
line of sight to the Vela SNR, was observed with ASCA for 120 ks. 
We find that the X-ray spectrum is featureless, and well described by
a power law, extending to three the class of 
shell-type SNRs dominated by nonthermal X-ray emission. Like G347.3$-$0.5, 
this low-latitude remnant displays discrete regions of enhanced emission 
along the rim as well as faint nonthermal emission from the interior. 
We derive limits on the thermal content of the remnant emission, although the
presence of the Vela SNR compromises our ability to seriously constrain
a low temperature component. Limits placed on the amount of Sc-K emission
are compared with the expected flux based upon the reported $^{44}$Ti emission
from \snr. We also report on an unresolved X-ray source surrounded by diffuse 
emission near the center of the remnant. The properties of the source 
are not well determined, but appear consistent with the interpretation
that the source is a neutron star surrounded by a synchrotron nebula.
Alternatively, the source may be associated with one of two stars located
within the positional error circle, but this appears somewhat unlikely.

\end{abstract}


\section{INTRODUCTION}

Young supernova remnants (SNRs) are believed to be the prime accelerators
of cosmic rays, at least up to the ``knee'' of the cosmic ray spectrum
at energies of $\sim 10^3$~TeV, beyond which
the spectrum steepens. However, the direct evidence for particle
acceleration to such energies is scarce. Radio emission from shell-type
SNRs is the result of synchrotron radiation from shock accelerated electrons,
but these electrons have typical energies of order 10~GeV. It is only
the recent discoveries of nonthermal X-ray emission from shell-type SNRs 
that have finally provided evidence of particles at energies of
10--100~TeV. The X-ray emission from SN~1006 (Koyama et al. 1995) 
and G347.3$-$0.5 (Koyama et al. 1997, Slane et al. 1999) is {\it
predominantly} nonthermal, while that from Cas~A, Kepler, Tycho,
and RCW 86 (Allen, Gotthelf, \& Petre 1999) is predominantly thermal,
but contains a nonthermal component as well. 
However, it may also be the X-ray emission that poses the most
significant constraint on accepting SNRs as the source of cosmic rays
at energies near the knee. The X-ray fluxes from all remnants fall
below the extrapolation of their radio spectra, implying a steepening
or cut-off of the electron spectrum at higher energies. The implied spectral
cut-off energies appear to be well below the knee of the cosmic ray
spectrum (Reynolds \& Keohane 1999). It is thus of considerable
importance to identify new examples of shell-type SNRs for which the
X-ray emission is largely nonthermal.

RX~J0852.0-0462 was discovered by Aschenbach (1998) using
data from the ROSAT All-Sky Survey. Situated along the line of sight
to the Vela SNR, the emission stands out above the soft
thermal emission from Vela only at energies above $\sim 1$~keV. Data 
from the ROSAT PSPC indicates a hard spectrum with either a hot
thermal component with $kT \sim 2.5$~keV or a power law component
with photon index $\Gamma \sim 2.6$ (Aschenbach 1998). The remnant
is circular with a diameter of $\sim 2^\circ$; the emission is 
particularly enhanced along the northern, western, and southeastern
limbs. Hereafter, we use the designation \snr\ for this newly
identified SNR.

As first noted by Aschenbach (1998) the radio emission
from \snr\ is quite weak ($S_\nu = 30 - 50$ Jy at 1 GHz) with faint,
limb-brightened emission similar to the X-ray morphology 
(Combi, Romero, \& Benaglia 1999; Duncan \& Green 2000). SN1006 
and G347 are also
faint radio emitters, perhaps suggesting a general characteristic
of remnants for which the X-ray flux is dominated by nonthermal
processes.

Iyudin et al. (1998) reported the COMPTEL detection of $^{44}$Ti from 
the source GRO~J0852-4642, which was tentatively associated with \snr.
If correct, this association would be profound; the very short $^{44}$Ti
lifetime ($\tau \approx 90$~y) would imply a very young SNR, and the
large observed size would require that the remnant be very nearby as
well. Estimates based on the X-ray diameter and $\gamma$-ray flux
of $^{44}$Ti indicate an age of $\sim 680$~y and a distance of
$\sim 200$~pc (Aschenbach, Iyudin, \& Sch\"{o}nfelder 1999).
The hard X-ray spectrum would seem to support this scenario
as well. The inferred temperature would imply a rapid shock that is
consistent with a young SNR. However, as we show here, the hard
X-ray emission is not from hot, shock-heated gas; it is nonthermal.
Further, separate reanalysis of the COMPTEL data finds that the detection of
\snr\ as a $^{44}$Ti source is only significant at the
$2 - 4 \sigma$ confidence level (Sch\"onfelder et al.~2000).
In the absence of such emission, and given that the X-ray emission
is nonthermal, the nearby distance and young age may need to be reexamined.

%%%%%%%%%%%%%%%% Figure 1 %%%%%%%%%%%%%%%%%%%%%%%%%%%%%%%%%%%%%%%%%
\centerline{\psfig{file=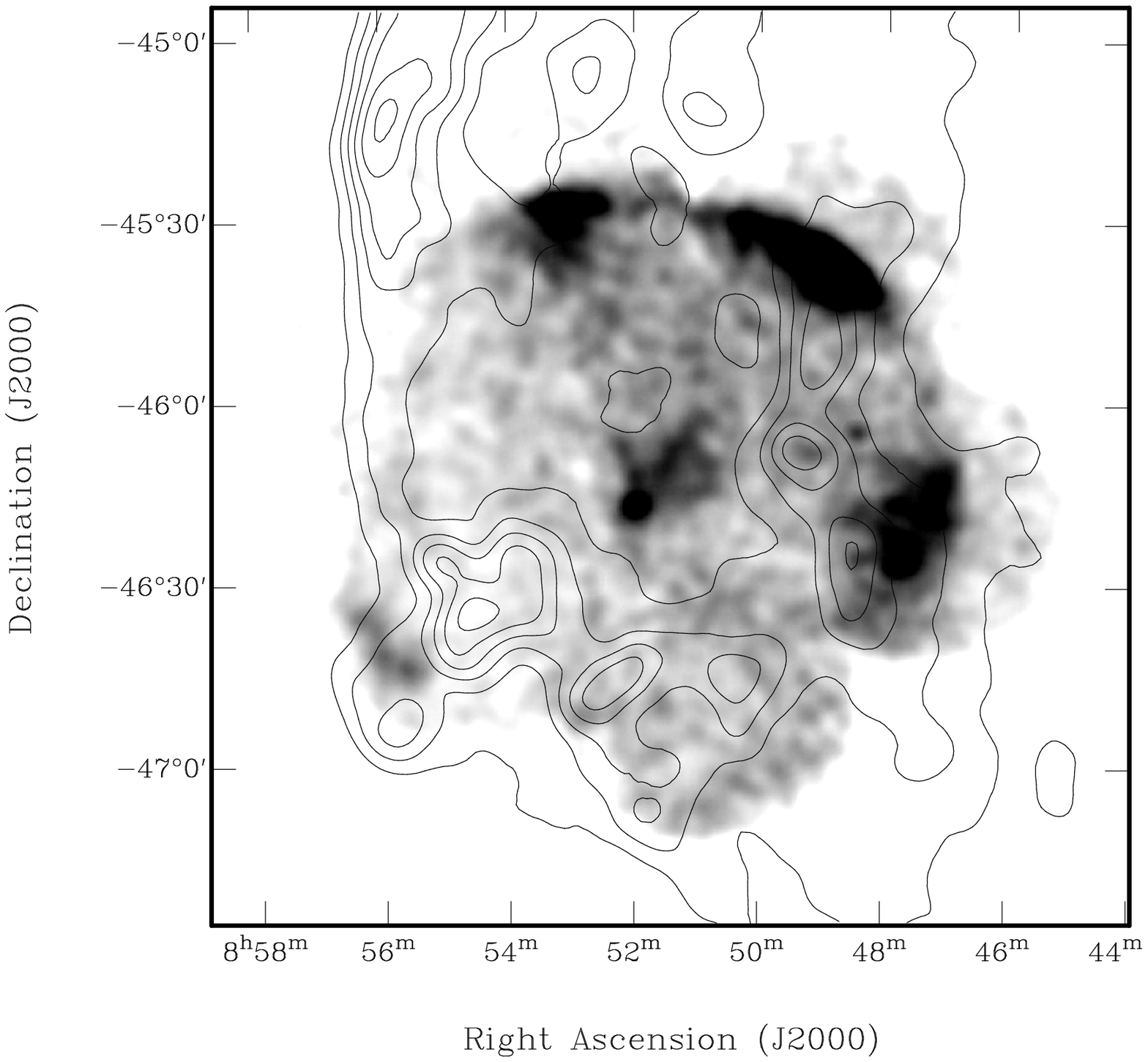,width=8.8cm}}
\small\parindent=3.5mm
{\sc Fig.}~1.---
ASCA GIS image of G266.2--1.2 ($E = 0.7 - 10$~keV). The image consists of
a mosaic of 7 individual fields. Contours represent the outline of the
Vela SNR as seen in ROSAT survey data with the PSPC.
\normalsize
%%%%%%%%%%%%%%%%%%%%%%%%%%%%%%%%%%%%%%%%%%%%%%%%%%%%%%%%%%%%%%%%%%%%

\section{OBSERVATIONS AND ANALYSIS}
We have carried out X-ray observations of \snr\ with the {\it Advanced
Satellite for Cosmology and Astrophysics} (ASCA). The remnant was
mapped in 7 distinct pointings, each of $\sim 17$~ks duration. The
resulting image from the ASCA GIS detectors is illustrated in 
Figure 1 (see also Tsunemi et al.~2000).
Bright emission along the northern and western shell is accompanied by a 
compact central source surrounded by diffuse emission. Additional enhanced
emission along the southeast shell is seen in the ROSAT image (Aschenbach
1998) but was inadvertently missed in the mapping carried out here.
Contours in Figure 1 correspond to soft emission ($E = 0.1 - 2.4$~keV)
from the Vela SNR, as observed in the ROSAT All-Sky Survey.
Standard screening processes were applied to data from both the GIS and
SIS detectors, and spectra were extracted from distinct regions of the
remnant. Regions outside the SNR shell in the northwestern pointing were
used for background subtraction for the GIS and for SIS chips S0C1,2 and
S1C0,3. For the other SIS chips, for which no local background was 
available, we extracted background spectra from the blank sky fields
available from the ASCA Guest Observer Facility. Although these fields
are at high Galactic latitude, and are thus not representative of our
viewing direction (particularly given the contribution from Vela
itself), we find that results from spectral fitting 
are quite similar with either background source. 

In Figure 2 we present plots of the X-ray spectra from three regions
along the rim of the remnant using the GIS detectors: 1) the bright 
northwest rim; 2) the northeast rim; and 3) the western rim.
Spectra were extracted from circular regions with radius $\sim 10$~arcmin
centered on the location of maximum brightness.
Unlike the line-dominated thermal emission
that one expects from a young remnant, the spectra are featureless
and well described by a power law of index $\sim 2.6$. This is remarkably
similar to that observed for G347.3$-$0.5 (Koyama et al.~1997,
Slane et al.~1999), another
shell-type SNR dominated by nonthermal emission. The index is flatter
than that observed for SN~1006 ($\Gamma = 2.95 \pm 0.2$), the prototype of this 
class of nonthermal shell-type SNRs (Koyama et al.~1995), although this
value was derived for only a portion of the remnant shell.

Due to the smaller field of view and the difficulty of merging
spectra from different CCDs, the spectra from the SIS are somewhat sparse.
However, there is still a clear lack of emission line features.
The spectra are still
adequately described by an absorbed power law, although the best-fit
parameters differ somewhat from those derived from the GIS observations.
Due to the better overall statistics, and the higher sensitivity to
the high energy end of the spectrum, we use the results from the GIS
spectral fits in the discussion below.

Although we have applied a background subtraction to the GIS spectra, the
soft thermal emission from the Vela SNR is spatially variable, and
there is significant residual soft flux in some of the regions, as
seen in Figure 2. We have modeled the emission with two components.
A thermal model (Raymond \& Smith 1977) was used to account for the
soft emission, with the column density fixed at $\sim 10^{20}
{\rm\ cm}^{-2}$, a typical value for the Vela SNR (Bocchino,
Maggio, \& Sciortino 1999; Aschenbach, Egger, \& Tr\"umper 1985). 
An absorbed power law was used for the
hard emission. To emphasize the contribution from Vela, we have not
included the thermal component in the spectral plots shown here.

The best-fit spectral parameters for \snr\ are summarized 
in Table 1. The temperature of the thermal component is similar to
that observed for the hotter component of Vela (Bocchino, Maggio, 
\& Sciortino 1999), as expected since the GIS response is not well-suited
to detected the softer component at $\sim 0.1$~keV. The spectral
indices for the emission regions along the shell of \snr\ are
all consistent with a value of $\sim 2.6.$

%%%%%%%%%%%%%%%55%%%%%%%% Figure 2 %%%%%%%%%%%%%%%%%%%%%%%%%%%%%%%%%%
\vspace{-0.05in}
\centerline{\psfig{file=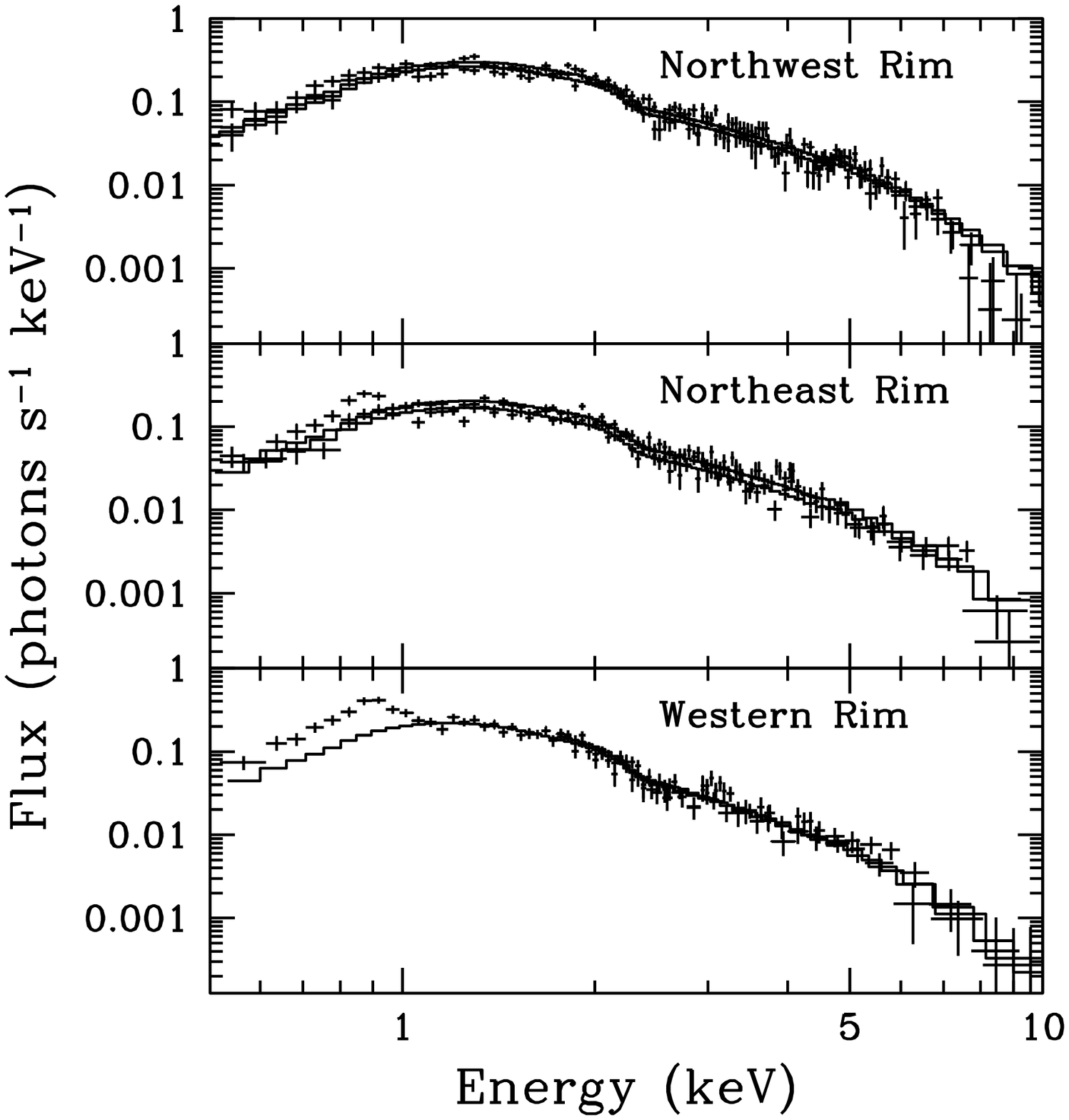,width=8.8cm}}
\small\parindent=3.5mm
{\sc Fig.}~2.---
ASCA spectra from both GIS detectors for regions of G266.2--1.2.
The featureless spectrum
is well described by a power law. Excess flux at low energies is
presumably associated with soft thermal emission from the Vela SNR.
\normalsize
%%%%%%%%%%%%%%%%%%%%%%%%%%%%%%%%%%%%%%%%%%%%%%%%%%%%%%%%%%%%%%%%%%%%%

The column density for the power law component summarized in Table 1 is
significantly higher than that for Vela. While simple scaling of
the column density to estimate the distance to \snr\ is clearly
rather uncertain, it would appear that the remnant is at least
several times more distant than Vela. 
To provide some measure of the scaling of quantities, in discussions 
below, we express the distance as $d_1 = (d/1{\rm\ kpc})$. 
We note that the column density derived from PSPC data alone (Aschenbach
1998), though not well determined, is apparently lower than that
measured here. Given
the importance of the distance measurement, as we discuss below, it
is of considerable interest to obtain improved measurements of the
column density.

\begin{center}
\small
{\bf Table 1: G266.2--1.2: Spectral Parameters}\\
\begin{tabular}{l|llll}\hline\hline
Region & $kT$(keV) & $N_H^a$ & $\Gamma$ (photon) & $F_x^{b}$\\ \hline
NW Rim & $0.5^{+0.2}_{-0.3}$ & $4.0 \pm 1.8$ & $2.6 \pm 0.2$ & $4.2 \times
10^{-
11}$ \\ %updated 4/19
NE Rim & $0.6 \pm 0.1$ & $5.3^{+2.0}_{-0.9}$ & $2.6 \pm 0.2$ & $2.9 \times
10^{-
11}$\\ %updated 4/19
W Rim & $0.5^{+0.1}_{-0.2}$ & $1.4^{+2.8}_{-1.4}$ & $2.5 \pm 0.2$ & $2.1
\times 10^{-11}$ \\ %updated 4/19
Center$^c$ & $0.7 \pm 0.1$ & $11.5^{+14.3}_{-10.2}$ & $2.0^{+0.6}_{-0.3}$ &
$6.7 \times 10^{-12}$ \\ \hline %updated 4/19
\multicolumn{5}{c}{AX J0851.9-4617.4} \\ \hline
BB & $0.47 \pm 0.04$ & $0.0 - 0.8$ & & $2.1  \times
10^{-12}$ \\
PL & & $3.7^{+3.8}_{-2.6}$ & $3.2 \pm 0.5$ & $6.6 \times
10^{-12}$ \\ \hline
\end{tabular}

$a)$ Power Law component only for SNR, in units of $10^{21}{\rm\ cm}^{-2}$\\
$b) {\rm\ erg\ cm}^{-2} {\rm\ s}^{-1}$ (0.5 - 10 keV) \\
$c)$ diffuse emission only
\end{center}
\normalsize

As shown in Figure 1, the central region of the remnant contains a
compact source which we designate as AX~J0851.9-4617.4.
The image centroid is at RA$_{2000}$: 08$^h$51$^m$57$^s$,
Dec$_{2000}$: $-46^\circ$17$^\prime$24$^{\prime\prime}$; 
the position uncertainty is
roughly 2~arcmin in each direction due to the low-resolution mode
in which the GIS data were obtained.
This source is surrounded by diffuse emission that
extends toward the northwest. It is of considerable interest to speculate
as to whether this represents a central neutron star surrounded by
a synchrotron nebula. Such an association would be spectacular indeed,
providing further evidence of a massive progenitor for the remnant,
and possibly constraining the age through modeling of the plerionic
component. In Figure 3 we present the GIS spectrum for both AX~J0851.9-4617.4
and the diffuse central region of the remnant. The diffuse emission is
well described by a power law of spectral index $\sim 2.0$, again accompanied
by a soft thermal component that we associate with Vela. Given the
nonthermal nature of the shell emission, however, it is quite possible
that the diffuse central emission is just associated with emission from the
shell projected along the central line of sight. We note that the spectrum
of this central emission appears somewhat harder than that from the rest of
the remnant, perhaps suggesting a plerionic nature, but more sensitive
observations are required to clarify this.

The GIS count rate for AX~J0851.9-4617.4 is $\sim 0.04 {\rm\ s}^{-1}$.
The $14.6$~ks GIS exposure (after screening) thus yields roughly
500 counts in each detector. The spectrum is thus rather sparse and can
be adequately described by a variety of models. Two of particular interest are
blackbody emission, as might be expected from the surface of a cooling
neutron star or from the polar caps of a neutron star heated through particle
acceleration in the magnetosphere, and a power law spectrum that might result
from emission directly from a neutron star magnetosphere.  Best-fit spectral
results from these models are summarized in Table 1.

%%%%%%%%%%%%%%%%%%%%%%%%%% Figure 3 %%%%%%%%%%%%%%%%%%%%%%%%%%%%%%%%
\centerline{\psfig{file=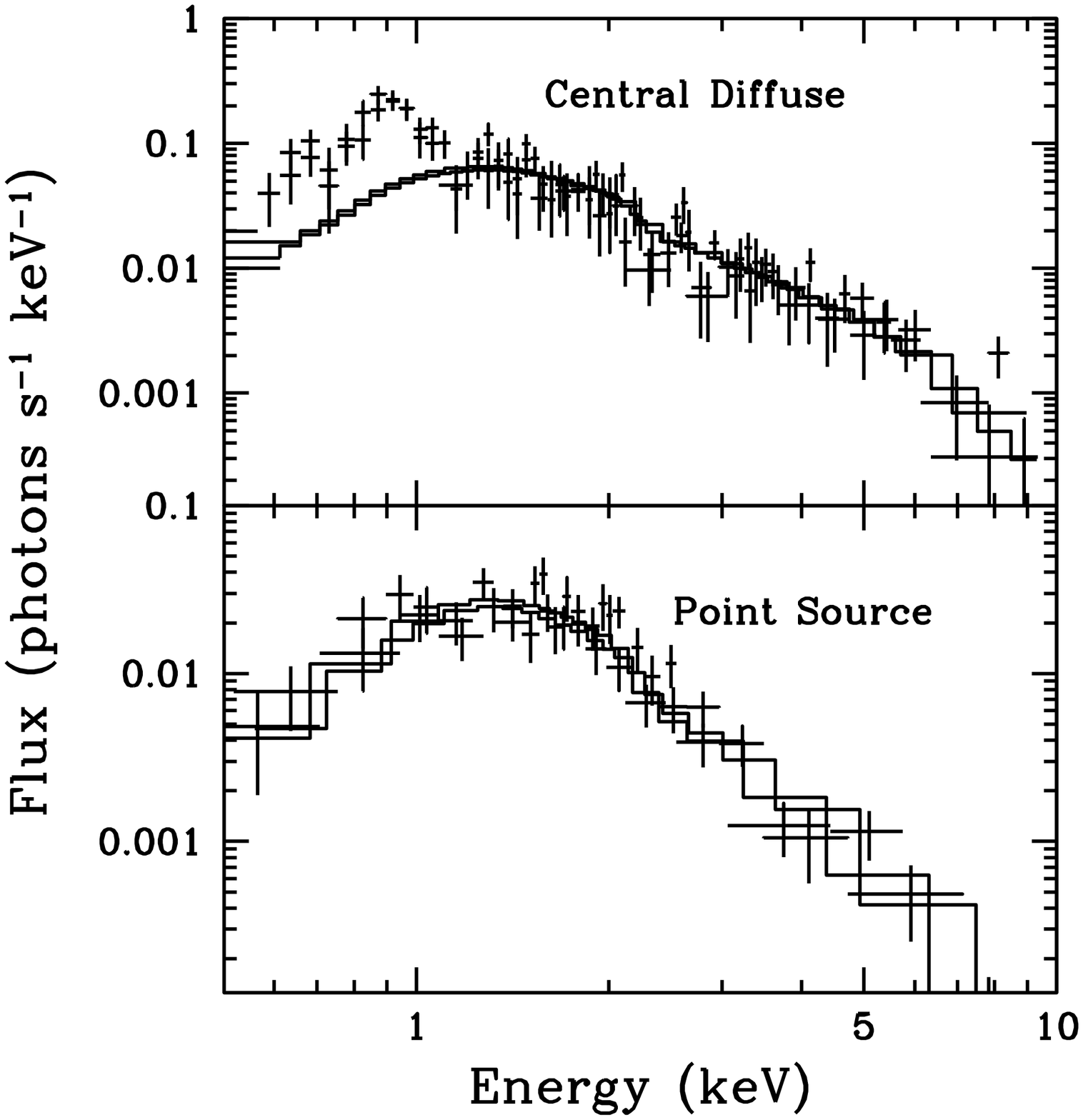,width=8.8cm}}
\small\parindent=3.5mm
{\sc Fig.}~3.---
ASCA spectra from both GIS detectors for the central point source and
surrounding diffuse region
of G266.2--1.2. The diffuse emission is indistinguishable from that
of the SNR shell. The point source spectrum is adequately described by
a power law, but other models are possible as well.
\normalsize
%%%%%%%%%%%%%%%%%%%%%%%%%%%%%%%%%%%%%%%%%%%%%%%%%%%%%%%%%%%%%%%%%%%%%

\section{DISCUSSION} 
The prevalence of nonthermal X-ray emission in \snr\ suggests that the
remnant is an efficient accelerator of particles. While thermal emission
from swept-up material and shocked ejecta typically dominates the
emission of shell-type SNRs, there are conditions which can lead to 
enhanced nonthermal X-ray emission. Recent models for nonlinear shock
acceleration in SNRs (e.g. Ellison, Berezhko, \& Baring 2000) show that
the ambient density and magnetic field are critical parameters in
determining the relative contributions of thermal and nonthermal electrons
in the X-ray emission. Low values of both quantities can yield
a synchrotron-dominated remnant in X-rays,  while a low magnetic field,
in particular, also leads to an increased ratio of the TeV $\gamma$-ray to 
radio flux ratio. This scenario appears
consistent with both SN1006 and G347.3$-$0.5. The former is the remnant
of a Type Ia explosion, for which the ejecta mass and ambient density
are expected to be relatively low, and is located at a relatively high
Galactic latitude which is consistent with a low magnetic field. The
latter, on the other hand, is in the Galactic plane and appears consistent
with a massive star progenitor if the association with surrounding
molecular clouds and a nearby \ion{H}{2} region suggested by Slane
et al. (1999) is correct. This, too, would then imply a low ambient
density if the remnant has evolved in a stellar wind cavity, which
is likely in such a scenario. Both remnants are observed in TeV 
$\gamma$-rays (Tanimori et al. 1998, Muraishi et al. 2000), and both are 
weak radio remnants (Stephenson et al. 1977, Aschenbach 1998,
Combi et al. 1999).
We note, however, that the circumstances which lead to a dominant
nonthermal contribution in the soft X-ray band are complicated. Reynolds
(1998) has shown that the maximum nonthermal X-ray emission relative
to the radio emission occurs when the remnant age is comparable to
the electron loss time in the postshock magnetic field. However, this
alone is not sufficient to guarantee that the thermal X-ray emission
from shock-heated ejecta and interstellar material will not dominate
the X-ray flux. 

For \snr, the ASCA observations clearly show that nonthermal processes
dominate the X-ray emission. Our limit on the density of any thermally
emitting material in the northwest region, where the emission is brightest,
is $n_0 < 2.9 \times 10^{-2} d_1^{-1/2} f^{-1/2} {\rm\ cm}^{-3}$, where $f$
is the filling factor of a sphere taken as the emitting volume in
the region extracted. Here we have assumed an equilibrium thermal model
(Raymond \& Smith 1977). Nonequilibrium ionization models can yield higher
emissivities, which would lower the derived limit on the density.
We note that this limit is restricted to
gas with temperatures above 1~keV because of the severe contributions of 
soft thermal emission from the Vela SNR; higher densities of cooler material
in \snr\ cannot be ruled out. Observations of the brightest shell regions
with high angular resolution, so as to minimize these contributions, are
of considerable interest to further constrain the ratio of thermal to
nonthermal emission.

If \snr\ is actually a source of $^{44}$Ti at levels indicated by the
COMPTEL measurements, so that the age and distance
estimates suggested by Aschenbach et al. (1999) and
others are correct, then the ambient density is extremely low. The
reported $\gamma$-ray line flux would indicate a massive star progenitor,
and the low density would then imply the presence of a stellar-wind
cavity such as that inferred for G347.3$-$0.5. 

%%%%%%%%%%%%%%%%%%%%%%%%%%% Figure 4 %%%%%%%%%%%%%%%%%%%%%%%%%%%%
\centerline{\psfig{file=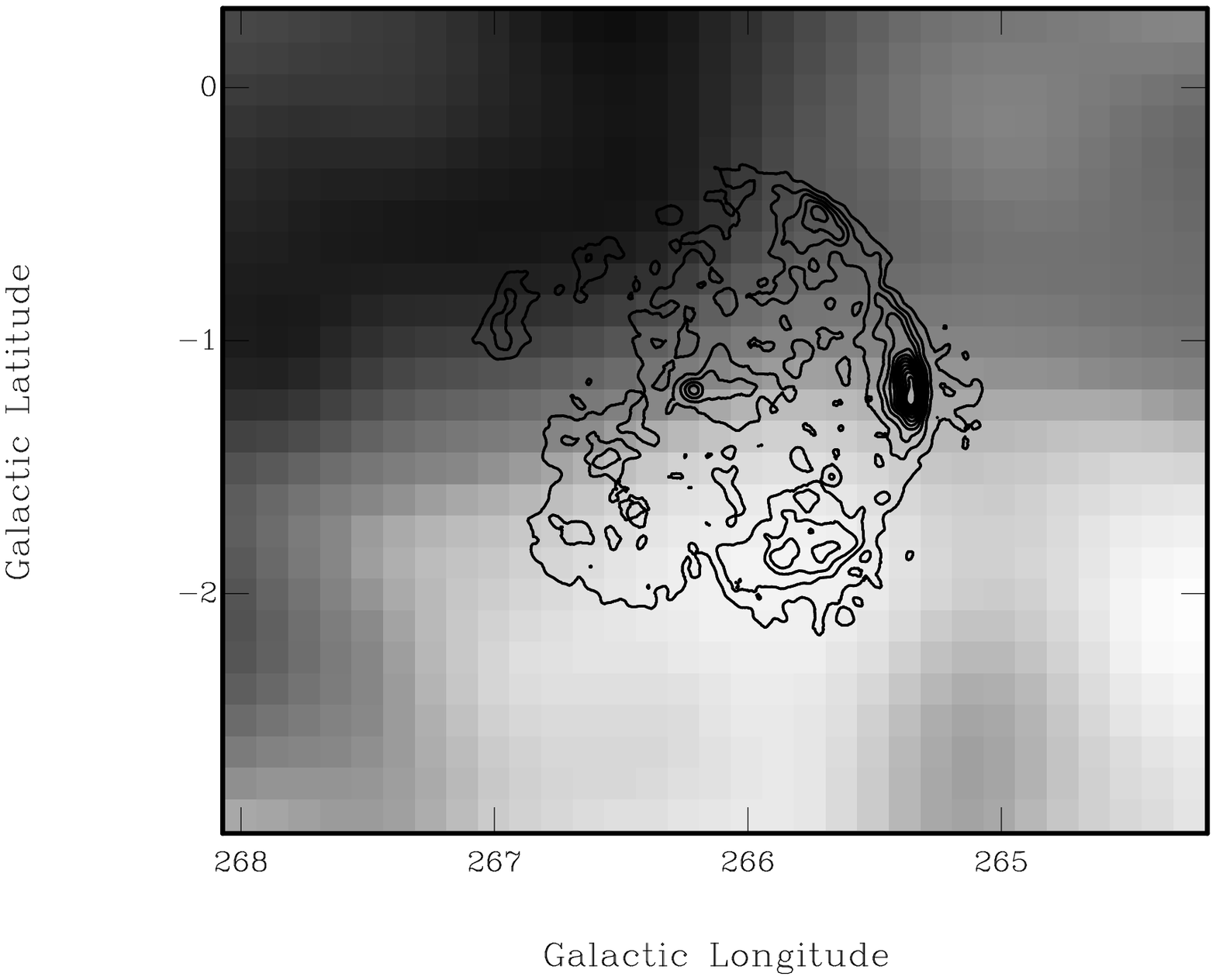,width=9.2cm}}
\small\parindent=3.5mm
{\sc Fig.}~4.---
CO emission ($V_{LSR} = -5 - +20 {\rm\ km\ s}^{-1}$)
along the line of sight to \snr. Countours correspond to the X-ray
emission from the remnant as seen in Figure 1. Note the use of
Galactic coordinates to emphasize orientation relative to the
Galactic Plane.
\normalsize
\vspace{0.1in}
%%%%%%%%%%%%%%%%%%%%%%%%%%%%%%%%%%%%%%%%%%%%%%%%%%%%%%%%%%%%%%%%%%

CO data (May, Murphy, \& Thaddeus 1988) reveal a concentration of
giant molecular clouds -- the Vela Molecular Ridge -- at a distance 
of $\sim 1 - 2$~kpc in the direction of Vela. The presence of
OB and R associations, as well as \Hii\ regions, demonstrate that
these clouds are sites of recent star formation. The bulk of the
CO emission along the line of sight to \snr\ is concentrated in a
velocity range $v_{\rm LSR} = 0.7 - 9.8 {\rm\ km\ s}^{-1}$ (Murphy
\& May 1991), and extends down to a Galactic latitude of 
$b \approx -1.2^\circ$. The total column density through the ridge
is in excess of $10^{22}{\rm\ cm}^{-2}$, with the northeast rim
of \snr\ (nearest the Galactic Plane) falling along the steeply 
increasing column density region
of the ridge while the western rim lies along a line of much lower
column density (Figure 4); the CO column density varies by more
than a factor of 6 between these regions. The lack of a strong variation 
in $N_H$ across
\snr\ indicates that the remnant cannot be more distant than the
Vela Molecular Ridge. On the other hand, if $N_H$ is
much larger than for the Vela SNR, as the GIS data reported here
indicate, then \snr\ must be as distant as possible consistent with
being in front of most of the Molecular Ridge gas.
This is consistent with an interpretation in
which \snr\ is the product of a massive progenitor from this star-forming
region, and that the remnant evolution has proceeded in the presence
of a wind cavity, leading to a low ambient density much like that
indicated for G347.3$-$0.5.

The angular size of \snr\ gives a free-expansion age of
$t \sim 3.4 \times 10^3 v_{\rm 5k}^{-1} d_1$~yr, where $v_{\rm 5k}$ is
the expansion velocity in units of $5000 {\rm\ km\ s}^{-1}$. 
Alternatively, using the limit on the preshock density derived above,
the Sedov age of the remnant is $t < 3.9 \times 10^3 d_1^{9/4}
f^{-1/4} E_{51}^{-1/2}$~yr, where $E_{51}$ is the kinetic energy from the
explosion in units of $10^{51}$~erg. In either case, the remnant
is relatively young, although ages as young as those based on the 
$^{44}$Ti lifetime would require either an extremely large expansion
velocity or an extremely low ambient density unless the remnant
distance is actually much less than 1 kpc.

We note that the $E = 1.156$~MeV $\gamma$-rays that announce the
presence of $^{44}$Ti are actually secondary products. The decay of
$^{44}$Ti proceeds via electron capture to form $^{44}$Sc which then
decays rapidly to $^{44}$Ca, producing the observed $\gamma$-rays.
As electrons fill the vacancy created by the electron capture, we
expect Sc-K emission of X-rays at $E = 4.1$~keV. Using the fluorescence
efficiency of $19\%$ (Krause 1979), we expect a flux $F({\rm Sc-K}) \sim
(6.6 \pm 1.3) \times 10^{-6}{\rm\ cm}^{-2}{\rm\ s}^{-1}$ 
based on the $^{44}$Ti flux
of Iyudin et al. 1998). We have searched the ASCA SIS data for evidence
of such a line feature, but have found no compelling evidence to support
its presence. The SIS0 spectrum of one region along the NW rim contains
a feature at $\sim 4$~keV (Figure 5), 
but is not seen in the SIS1 spectrum of the same region,
nor in adjacent regions with either detector. 
Using our power law fits, the inclusion of the additional feature
is not significant according to the F-test; 
the thermal fits of Tsunemi et al. (2000) suggest that the feature
is significant at the $\sim 99\%$ confidence level.
We derive a $1 \sigma$ 
upper limit of  $4.4 \times 10^{-6}{\rm\ cm}^{-2}{\rm\ s}^{-1}$ for
Sc-K emission from this region, which is not inconsistent with the 
expected result, particularly since additional contributions could
come from elsewhere in the remnant.
Higher sensitivity searches for such emission are clearly of interest
to provide an independent method for obtaining evidence of the presence
of $^{44}$Ti.

The nature of the central compact X-ray source in \snr\ is poorly constrained
by the ASCA and ROSAT data. There are two bright stars ($m_V < 14$)
located within the large 
positional uncertainty (Aschenbach 1998), but neither is a compelling 
counterpart. 
The first star, WRAY~16-30 (IRAS 08502-4606), is a Be star whose position
is within 2~arcmin of AX~J0851.9-4617.4. Using a thermal model appropriate
for a Be star, the X-ray luminosity of the source is $L_x \sim 4 
\times 10^{32} d_1^2 {\rm\ erg\ s}^{-1}$. The typical range for 
Be stars is roughly $10^{29} - 10^{32} {\rm\ erg\ s}^{-1}$
(Zinnecker \& Preibisch 1994). This would suggest a distance
of less than $\sim 500$~pc, which is difficult to reconcile with the
observed visual magnitude of $V = 13.8.$ The second potential counterpart
is HD76060 ($m_V = 7.88$, $m_B = 7.79$), a B8 IV/V star whose measured
parallax puts the distance at $\sim 270$~pc. The upper end of the
X-ray luminosity for stars of this class is $L_X \approx 10^{30.5}
{\rm\ erg\ cm}^{-2}{\rm\ s}^{-1}$ (Bergh\"ofer, Schmitt, Danner,
\& Cassinelli 1997) which, for the measured value, requires a distance
of roughly 90~pc. While this discrepancy is perhaps not sufficient to
absolutely rule out an association, further observations to improve the
position of the X-ray source, and better constrain its spectrum, are
clearly in order. 

%%%%%%%%%%%%%%%%%%%%%%%%% Figure 5 %%%%%%%%%%%%%%%%%%%%%%%%%%%%%%%5
\centerline{\psfig{file=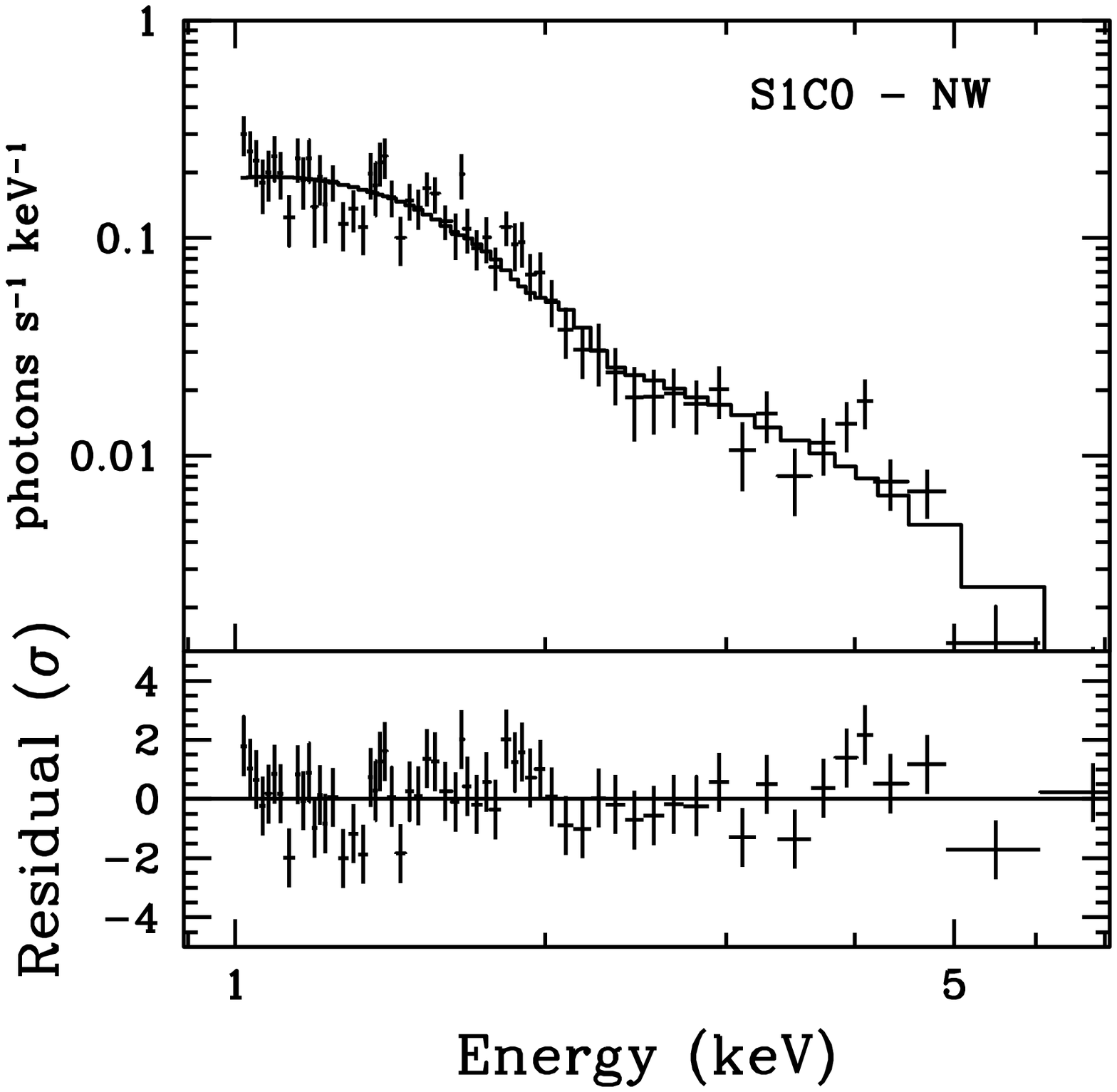,width=9cm}}
\vspace{-0.3in}
\small\parindent=3.5mm
{\sc Fig.}~5.---
ASCA SIS spectrum (chip S1C0) for a region along the NW limb. There
is some evidence of a feature at $E \sim 4$~keV, perhaps suggestive
of $^{44}$Sc emission. However, this feature is not seen in other
SIS data.
\normalsize
\vspace{0.1in}
%%%%%%%%%%%%%%%%%%%%%%%%%%%%%%%%%%%%%%%%%%%%%%%%%%%%%%%%%%%%%%%%%%%%%%%%

Alternatively, we have considered the possibility that AX~J0851.9-4617.4 is 
a neutron star associated with \snr. In this case, the power-law fit
yields a spectral index that is rather steep compared with known pulsars.
The luminosity, on the other hand, is quite reasonable for a young neutron
star if the distance is indeed several kpc, and the observed column density is
compatible with that for the SNR shell. A blackbody model leads to an
inferred surface temperature of $\sim 0.5$~keV with an emitting region
only $\sim 200  d_1 {\rm\ m}$ in radius which could be indicative of
emission from compact polar cap regions of a neutron star, although this
region is somewhat large for such a scenario. We have searched
for pulsations but detect no obvious modulation of the flux. We set an
upper limit to the pulsed fraction of $45 - 70\%$ for frequencies from
$10^{-3} - 10^3$~Hz.

Finally, we note that the spectrum of the diffuse emission in the central
region of the remnant is consistent with a power law whose index is
similar to the Crab nebula and other synchrotron nebulae associated
with supernova remnants. The ratio of the point source flux to that
of the diffuse surrounding emission is $\sim 40\%$ which is similar
to that for the Vela pulsar and its associated synchrotron nebula.
It is of interest that the Vela pulsar also has a soft spectrum
best characterized as blackbody emission from heated polar caps
(\"{O}gelman, Finley, \& Zimmermann 1993). 
High resolution radio observations of the central region of \snr\ 
using the ATCA reveal a faint, slightly extended ($\sim 45''$ across)
source within the error circle of AX~J0851.9-4617.4 (B. Gaensler -- 
private communication). The 1.4~GHz flux density of this source
is $\sim2$~mJy, however, substantially lower than seen for known
synchrotron nebulae, and the X-ray luminosity is quite low as well. Further
observations are required to better understand the nature of this
source.

\section{CONCLUSIONS} 

The ASCA observations of \snr\ reveal that the soft X-ray emission from
this SNR is dominated by nonthermal processes. This brings to three the
number of SNRs in this class, and provides additional evidence for
shock acceleration of cosmic rays in SNRs. Because of the bright and
spatially varying background caused by the Vela SNR, limits on the
thermal emission from \snr\ are difficult to establish. As a result,
its evolutionary state is poorly constrained. However, the ASCA data 
reveal a larger column density for this remnant than for Vela,
indicating that \snr\ is at a larger distance,
and perhaps associated with the star formation region located at a
distance of $1 - 2$~kpc.
The very strong ratio of nonthermal to thermal X-ray emission argues for
a low density environment, possibly suggesting evolution in a stellar
wind cavity. Such a scenario would indicate a massive progenitor, from
which one could expect a relic neutron star. The point source 
AX~J0851.9-4617.4 could represent this source, with the diffuse emission
around the source being an associated synchrotron nebula, albeit
a very faint one. Higher resolution
X-ray observations of this source are of considerable importance in
order to assess its relationship to the SNR.

We note that TeV $\gamma$-ray emission has been detected from both
SN~1006 and G347.3$-$0.5, confirming the presence of very energetic
electrons in these remnants. Similar observations of \snr\ are clearly
of considerable interest as well.

\acknowledgments 
The authors would like to thank Roland Diehl and Donald Ellison for 
particularly helpful discussions in the preparation of this paper, and
also Bryan Gaensler, Douglas Bock, and Anne Green for kindly providing
information on radio measurements of AX~J0851.9-4617.4, and Tom Dame
for providing the CO data.
This work was supported in part by the National Aeronautics and
Space Administration through contract NAS8-39073 and grant NAG5-9106,
and has made use of data obtained from the High Energy Astrophysics
Science Archive Research Center (HEASARC), provided by NASA's Goddard
Space Flight Center.


\begin{references} 

\reference{} Allen, G. E., Gotthelf, E. V., Petre, R. 1999, 
``Evidence of 10-100 TeV Electrons in Supernova Remnants,'' in
Proceedings of the 26th International Cosmic Ray Conference, Salt Lake City,
17-25 August 1999, Edited by D. Kieda, M. Salamon, and B. Dingus,
Vol. 3, 480-3


\reference{} Aschenbach, B. 1998, Nature 396, 141  

\reference{} Aschenbach, B., Egger, R.,  \& Tr\"umper, J. 1995, Nature 373, 
587  

\reference{}Aschenbach, B.,  Iyudin, A. F., \& Sch\"{o}nfelder, V. 1999,
A\&A 350, 997

\reference{} Bergh\"ofer, T. W., Schmitt, J. H. M. M., Danner, R., \&
Cassinelli, J. P. 1997, A\&A 322, 167

\reference{} Bocchino, F., Maggio, A., \& Sciortino, S. 1999, A\&A 342, 839

\reference{} Combi, J. A., Romero, G. E., \& Benaglia, P. 1999, 
ApJL 519, L177

\reference{} Duncan, A. R., \& Green, D. A. 2000, A\&A - in press.

\reference{} Ellison, D. C., Berezhko, E. G., \& Baring, M. G. 2000, ApJ -
submitted

\reference{} Iyudin, A. F. et al. 1998, Nature 326, 142  

\reference{} Koyama, K., Petre, R., Gotthelf, E. V., Hwang, U., Matsura, M., 
Ozaki, M., \& Holt, S. S. 1995 Nature 378, 255  

\reference{} Koyama, K. et al.~1997, PASJ 49, L7  

\reference{} Krause, M. O. 1979, J. Phys. Chem. Ref. Data, 8, 307

\reference{} May, J., Murphy, D. C., \& Thaddeus, P. 1988, A\&ASS 73, 51

\reference{} Muraishi, H. et al. 2000, A\&A 354, L57

\reference{} Murphy, D. C., \& May, J. 1991, A\&A 247, 202

\reference{} \"{O}gelman, H., Finley, J. P., \& Zimmermann, H. U. 1993,
Nature 361, 136

\reference{} Raymond, J. C., \& Smith, B. W. 1977, ApJS, 35, 419

\reference{} Reynolds, S. P. 1998, ApJ, 493, 375

\reference{} Reynolds, S. P. \& Keohane, J. W. 1999, ApJ 525, 368

\reference{} Sedov, L. 1959, Similarity and Dimensional Methods in Mechanics 
(New York:Academic)

\reference{} Sch\"onfelder, V. et al. 2000, 5th Compton Symposium, Portsmouth,
AIP Conf. Proc. 510, p. 54, eds. M.L. McConnell and J.M. Ryan

\reference{} Slane, P., Gaensler, B. M., Dame, T. M., Hughes, J. P., Plucinsky,
P. P., \& Green, A., 1999, ApJ, 525, 357

\reference{} Stephenson, F. R., Clark, D. H., Crawford, D. F. 1977, MNRAS
180, 567

\reference{} Tanimori, T. et al. 1998, ApJ 497, L25

\reference{} Tsunemi, H., Miyata, E., Aschenbach, B., Hiraga, J., \&
Akutsu, D. 2000, PASJ, in press

\reference{} Zinnecker, H. \& Preibisch, Th. 1994, A\&A 292, 152

\end{references}
\end{document}